\documentclass[a4paper,10pt,conference]{IEEEtran}
\usepackage[T1]{fontenc} 
\usepackage[utf8]{inputenc} 
\usepackage{graphicx} 
\usepackage{hyperref} 
\usepackage{multirow} 
\usepackage{tabularx} 
\usepackage{color} 
\usepackage{textcomp} 
\usepackage{amsmath} 
\usepackage{amssymb} 
\usepackage{amsfonts} 
\usepackage{amsxtra} 
\usepackage{wasysym} 
\usepackage{isomath} 
\usepackage{mathtools} 
\usepackage{txfonts} 
\usepackage{upgreek} 
\usepackage{enumerate} 
\usepackage{tensor} 
\usepackage{pifont} 
\usepackage{soul} 
\usepackage{booktabs}
\usepackage{cite}

\usepackage[colorinlistoftodos,textwidth=10mm,textsize=tiny,obeyFinal]{todonotes}
   \newif\ifdraft
   \draftfalse
   \ifdraft
      \newcommand{\yp}[1]{\todo[color=orange!40]{#1}}
      \newcommand{\mxp}[1]{\todo[color=yellow]{#1}}
   \else
      \newcommand{\yp}[1]{}
      \newcommand{\mxp}[1]{}
   \fi

\begin{document}
% \title{Cross-Platform Deep Issue Classification}
\title{A Bug or a Suggestion? An Automatic Way to Label Issues}
\author{\IEEEauthorblockN{Yuxiang Zhu\IEEEauthorrefmark{1}, Minxue Pan\IEEEauthorrefmark{1}, Yu Pei\IEEEauthorrefmark{2} and Tian Zhang\IEEEauthorrefmark{1}} 
\IEEEauthorblockA{\IEEEauthorrefmark{1}State Key Laboratory for Novel Software Technology, Nanjing University, China}
\IEEEauthorblockA{\IEEEauthorrefmark{2}Department of Computing, The Hong Kong Polytechnic University, Hong Kong, China}
\IEEEauthorblockA{zyx@smail.nju.edu.cn, mxp@nju.edu.cn, csypei@comp.polyu.edu.hk, ztluck@nju.edu.cn}
}

\maketitle

\begin{abstract}
More and more users and developers are using Issue Tracking Systems (ITSs) to report issues, including bugs, feature requests, enhancement suggestions, etc. Different information, however, is gathered from users when issues are reported on different ITSs, which presents considerable challenges for issue classification tools to work effectively across the ITSs. Besides, bugs often take higher priority when it comes to classifying the issues, while existing approaches to issue classification seldom focus on distinguishing bugs and the other non-bug issues, leading to suboptimal accuracy in bug identification. 

In this paper, we propose a deep learning-based approach to automatically identify bug-reporting issues across various ITSs. The approach implements the k-NN algorithm to detect and correct misclassifications in data extracted from the ITSs, and trains an attention-based bi-directional long short-term memory (ABLSTM) network using a dataset of over 1.2 million labelled issues to identify bug reports. Experimental evaluation shows that our approach achieved an F-measure of 85.6\% in distinguishing bugs and other issues, significantly outperforming the other benchmark and state-of-the-art approaches examined in the experiment.

% More and more users and developers are using Issue Tracking Systems (ITSs) to report issues, including bug reports, feature requests, enhancement suggestions, etc. However, for platforms such as GitHub, few issues are clearly classified, forcing developers to handling those issues manually. Meanwhile, in traditional ITS like Jira and Bugzilla where each issue is submitted with a must-filled issue type, issues are often misclassified. Existing approaches aiming to classifying issues seldom focus on distinguishing bug and non-bug issues and are often subject to the misclassification problem, resulting in poor accuracy. In this paper, we propose a deep-learning-based approach to automatically classifying issues. To further improve accuracy, we adopt the k-NN algorithm to detect and correct misclassifications in the labelled dataset. After that, an attention-based Bi-LSTM is trained to classify issues using a dataset of over 1.2 million labelled issues. Evaluations show that our approach achieved 85.6\% of F-measure by ten-fold validation, which significantly outperforms the state-of-the-art approaches. Moreover, our approach enlightens a new way to handle issues from all common ITS, while most of the other relevant works are platform-specific.
\end{abstract}


\section{Introduction}\label{mark-I.}

User feedback is crucial in requirements engineering and software process management\cite{maguire2002user}, and how to automatically collect and analyze user feedback is a major task in these field. With the development of open source software (OSS) movement, issue reports are playing an essential role in the two-way communication among end users, contributing developers and core developers\cite{merten2015requirements,bissyande2013got,bertram2010communication}. 

Issue Tracking Systems (ITSs), commonly used in large projects, provide platforms to help developers collect, maintain, manage, and track issues like bugs, feature requests, improvement suggestions, and other user feedback~\cite{janak2009issue}. Jira is one of the most famous and widely used ITSs~\cite{ortu2015measuring}. Many open source projects and organizations, such as Hibernate\footnote{https://hibernate.atlassian.net/}, Jboss\footnote{https://jira.jboss.org/}, Spring\footnote{https://jira.spring.io/}, and the 
Apache Software Foundation\footnote{https://issues.apache.org/jira/}, use Jira to manage issues.
An example issue in the Jira ITS for project `Spring Boot' is shown in Figure 1. The issue has a title, a type, and other properties like priority, status, and resolution. Another popular ITS is the one used by the world's leading software development platform GitHub. Figure 2 shows the overview of three issues collected from the GitHub project ‘google/guava’\footnote{https://github.com/google/guava/issues}. As we can see from the figure, for each issue, the package it affects, its status, and its type are listed. For easy presentation, we refer to this ITS simply as the GitHub ITS in the rest of this paper. Other popular Issue Tracking Systems include, e.g., Bugzilla, Redmine, and Mantis.

\begin{figure}
\includegraphics[width=3.3in]{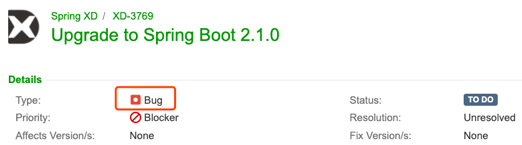}
\caption{An issue in the Jira ITS for project `Spring XD'.}
\label{fig:1}
\end{figure}

\begin{figure}
\includegraphics[width=3.3in]{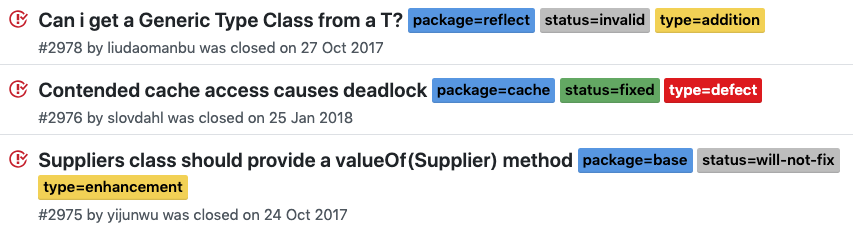}
\caption{Three issues in the GitHub ITS for project `google/guava'.}
\label{fig:2}
\end{figure}

In the open source community, a relatively higher priority is often placed on bug reports when it comes to triaging issues, since bug fixing is often more urgent than other tasks\cite{breu2010information}. Therefore, most ITSs allow reporters to manually label, or classify, an issue as reporting a bug or something else. Such classification, however, may be incorrect, due to limitations of the reporters' knowledge and experience. Such misclassifications may cause issues to be assigned to the wrong handler or delay the resolution of issues~\cite{bettenburg2008makes,wright2013estimating,kochhar2014s}. Besides, misclassifications constitute a major problem threatening the validity of models learned from ITSs~\cite{herzig2013s}, since they introduce noises to the learning process. 
For example, type is used to indicate whether an issue is a bug or not in Figure 1 and 2; The issue in Figure 1 is currently labeled as a bug, while it is more appropriate to be classified as a dependency-upgrade or enhancement request. A tool to accurately distinguish bugs from other types of issues can help developers better prioritize their tasks in processing the issues.

In this paper, we propose a novel approach to effective issue classification across various ITSs, where an attention-based bi-directional long short-term memory (ABLSTM) network is learned and used to distinguish bugs and non-bug issues. Particularly, the k-NN algorithm is employed in the approach To identify and correct misclassifications of issues extracted from existing ITSs, so that the model learned from these issues have greater discrimination power of bugs. Compared to existing state-of-the-art approaches and those based on traditional methods~\cite{antoniol2008bug,fan2017road,pingclasai2013classifying}, our approach significantly improves the F-measure in identifying bugs. 

Overall, we make the following contributions in this paper:

\begin{itemize}
\item We propose a novel approach to automatically and effectively labelling bug in various Issue Tracking Systems; % that can help reduce the amount of manual effort needed for bug identification;

\item We implement the approach into a prototype tool; 

\item We carry out an empirical study and show the superiority of our approach by comparing its effectiveness with existing state-of-the-art approaches and other benchmark approaches.

\end{itemize}

The remainder of this paper is structured as follows: Section~\ref{sec:methodology} presents the main steps involved in applying the proposed approach; Section~\ref{sec:experimental:design} reports on the experiments we conducted to evaluate the approach; Section~\ref{sec:results} gives the experimental results; Section~\ref{sec:threats} lists several major threats to our work's validity; Section~\ref{sec:relatedwork} reviews related work; Section~\ref{sec:conclusion} concludes the paper.

\section{Methodology}\label{sec:methodology}

In this section, we present our proposed approach, which can automatically label issues in ITSs. An overview of our approach is shown in Figure 3. The rest of this section describes the approach in detail. \yp{According to the figure, there is no other difference than the use of ``RNN Model Training'' between the training process and the testing process. Is this really the case?}

\begin{figure}
\centering
\includegraphics[width=3in]{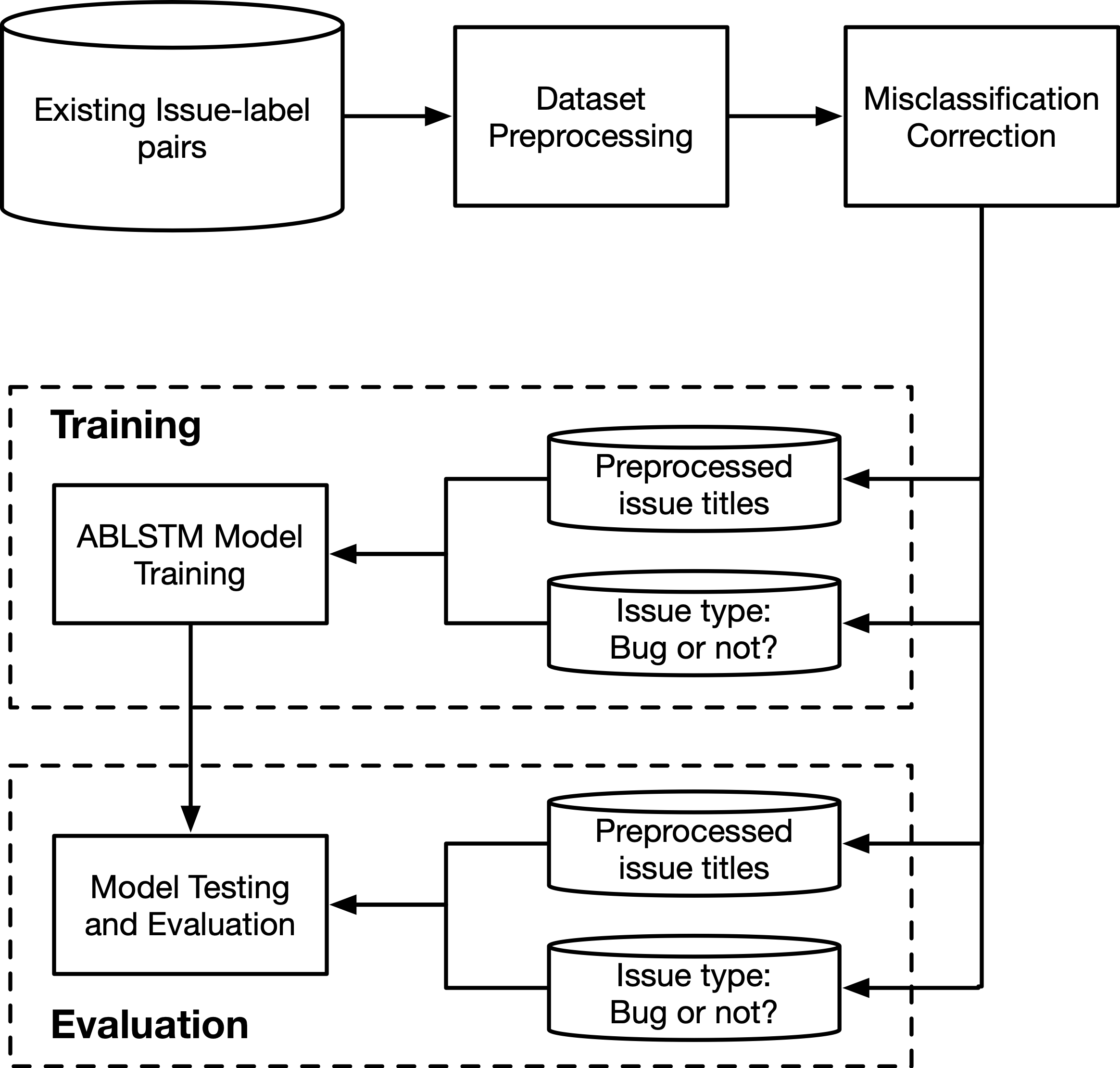}
\caption{The overall architecture of our approach}
\label{fig:3}
\end{figure}

\subsection{Data Collection}\label{sec:data:collection}

Our main data set is collected from three separate Jira ITSs, which respectively contains all issues in Apache, JBoss and Spring projects. The dataset contains over 1.2 million issues, all labelled by their reporters.

Moreover, we also collected issue data from GitHub. We noticed that only a few projects, even among high-stared projects, are comprehensively and carefully labelled. So we employed inclusion criteria to determine whether a project is suitable for data collection: 1) the number of issues exceeds 500; 2) more than half of the issues are labelled, suggesting the developers made serious attempt to manage issue labels; 3) labels are used to denote issues' types (some projects only use labels to denote the solving status, etc.). \yp{Criterion 3 requires that we know whether a label indicates a type or not already.}

After inspecting hundreds of top-stared Java projects in GitHub, 23 open source projects, where developers managed, labelled and solved issues cautiously, were discreetly selected. We collected 178390 issues from these projects in GitHub and 92.3\% of which are labelled. Among these labelled issues, 85833 issues are labelled with bug/non-bug information, while the other are only labelled with irrelevant information such as status and affected modules and will not be considered in the following processing. More details about our dataset are presented in Section~\ref{sec:dataset}.

\subsection{Dataset Preprocessing}\label{sec:reporter:type:extraction}

\textit{1) Reporter Type Extraction:} In different projects, reporters use different phraseology to denote the types of issues. For instance, labels like `bug', `bug report', `crash', and `defect' have been used to denote bugs, while labels like `suggestion', `enhancement', `improvement', and `enhancement request' have been used to denote non-bug issues. Besides, it is conventional in some projects to use prefixes like ``platform='', ``type='', and ``status='' to denote the labels' function, as demonstrated in Figure 2. 

To figure out the types of issues as assigned by their reporters, or the issues' \emph{reporter types}, we therefore extract the labels from all issues in our dataset, identify and remove the prefix in each label, and then decide manually whether the remaining of each label indicates a bug or a non-bug issue type. The reporter type of an issue is `bug' if at least one of its labels suggests so, `non-bug' if the issue is assigned with a type but other than `bug', or \emph{null} if no type is assigned to the issue by its reporter. 

%It is worth noting that 
Some labels provide other information relevant to issues, rather than their reporter types. For example, label `mac', and `pending', can be used to give the operating system where an issue occurred, and to indicate the processing status of an issue, respectively. These labels have no bearing on the reporter types of their related issues. 

Only issues with reporter type `bug' and `non-bug' are retained and used for further process (i.e., model training, as we describe in Section~\ref{sec:classifying:issues}), all other issues are discarded. 

\textit{2) Issue Title Preprocessing:} First, we stem all words from issue titles using NLTK WordNet Stemmer\cite{loper2002nltk}. Since we are going to compare similarities between sentences using the k-NN algorithm, it will be helpful to unify words in different tense and voices. Then, stop-words are filtered out as often done in natural language processing.

We notice that issue titles often contain names of program entities like Java classes, methods, and file names. Some of the names, e.g., `NullPointerException’ and `pom.xml’, are obviously useful for bug classification, but most of them, e.g., `JsonDecoder' and `derby.jar’, are too specific and not indicative of issue types. So we only maintain a dictionary of 20,000 most frequently occurred words and convert other words to token `{\textless}UKN{\textgreater}’. In this way, uncommon names are filtered out. Identifiers in camel case are not split into words in this step. 

Finally, we tokenize the issue titles using white space characters as delimiters and remove all punctuations.

\subsection{Misclassification Correction}\label{sec:misclassification:correction}

Next, we calculate the similarity between issues based on their titles and implement the k-NN algorithm to decide the actual types of retained issues.

\textit{1) Doc2Vec Training and Ball Tree Building:} We use Gensim\cite{rehurek2010software} doc2vec tools in our work to train document vector. Gensim.doc2vec is based on Distributed Memory (DM) by Quoc Le and Tomas Mikolov\cite{le2014distributed}. We traine a 128-dimension vector for every single issue in our dataset.

In k-NN algorithm, for every single issue, we have to search the entire data space to find its nearest neighbors, which is very time-consuming. Therefore, ball tree is introduced as a data structure to organize points in a multi-dimensional space\cite{omohundro1989five}. It can dramatically accelerate the search for nearest neighbors. We use the scikit-learn (Sklearn) toolkit\cite{pedregosa2011scikit}, a free machine learning library for Python, to build the ball tree model. %, which is used to improve searching speed in the next subsection.

\textit{2) Misclassification Correction:} In the traditional k-NN algorithm\cite{altman1992introduction}, an object is classified by a majority vote of its neighbors, which means even if the numbers of neighbors from two different classes are close, the algorithm will still give a result based on the narrow margin. Such narrow victory may result in uncertainty, contingency, and thus a decrease in the overall precision. 

Therefore, we enlarge this margin between different classification judgments. We predefine a judgment threshold $p$ (set to 0.8 by default), and only when the majority’s quantitative proportion equals or exceeds $p$, the object is classified to the majority. %Also we predefine k to 20 (as discussed in Section~\ref{sec:determine-kp}). 
In our misclassification identification process, for every single issue in our dataset, we first identify the $k$ nearest neighbors ($k$ is set to 20 by default). Then if at least $k*p$ of the neighbors are of a different type than the issue, we mark this issue as misclassified. Types of the misclassified issues are corrected at the end of the procedure. Correcting the type of an issue means changing the type from `bug' to `non-bug' or vice versa. %assigning the opposite relabelling the issue to its correct label. During correction, about 5\% of issues are corrected.

% To demonstrate this in detail, let's use a specific example: a single issue "installer log is missing" with type "non-bug". We first identified its 20 nearest neighbors. If 16 or more than 16 neighbors are 'bug', we will annotate this issue as 'misclassified' and at last correct it to type 'bug'. 

\subsection{Classifying Issues Using Neural Networks}\label{sec:classifying:issues}

% In this paper, we apply an attention-based bi-directional LSTM network (ABLSTM). It has been utilized in many fields, including relation classification\cite{zhou2016attention} and word sense disambiguation\cite{raganato2017neural}. Our model contains five component: input layer, embedding layer, LSTM layer, attention layer, and output layer. Figure 4 illustrates our network’s architecture.
In this paper, we apply an attention-based bi-directional LSTM (ABLSTM) network. Our model contains five components: input layer, embedding layer, LSTM layer, attention layer, and output layer. Figure 4 illustrates our network’s architecture.
 
 \begin{figure}
\includegraphics[width=3.3in]{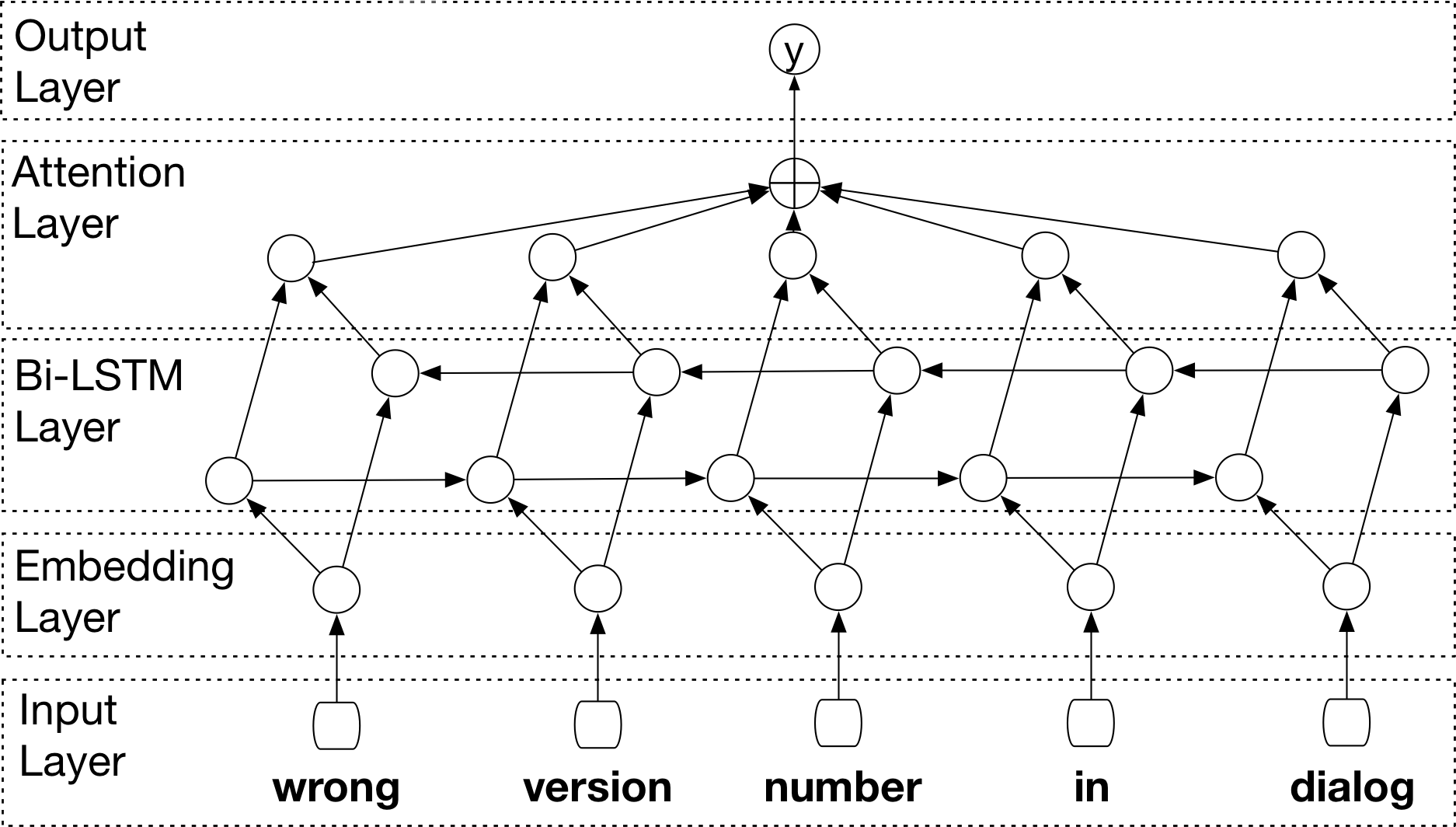}
\caption{The architecture of our neural network}
\label{fig:4}
\end{figure}

\textit{1) Embedding Layer:} In the embedding layer, every single token is mapped into a real-valued vector representation. Suppose $v_{i}$ is the one hot vector for single word $x_{i}$ in sentence \textit{S}, where $v_{i}$ has value 1 at index $c_{i}$ and 0 in any other positions. The embedding matrix \textit{M} is a parameter to be learned, and hyperparameters include the size of vocabulary \textit{V} and the dimension of word embedding \textit{d}. Then we can translate the word $x_{i}$ into its word embedding vector $e_{i}$ by calculating product of the one hot vector and the embedding matrix:
\begin{equation}
e_{i}=M\cdot v_{i} 
\end{equation}

Then, for sentence $S=\left\{x_{1},x_{2},x_{3},\ldots ,x_{n}\right\}$, we map each word to embedding vector correspondingly and then feed them into the next layer.

\textit{2) Bi-LSTM Layer:}

In order to solve gradient vanishing or exploding problem, LSTM introduces gate and memory mechanism to keep track of long time memory\cite{hochreiter1997long}. LSTM consists of a memory cell $c_{t}$ and three gates: input gate $i_{t}$, forget gate $f_{t}$, and output gate $o_{t}$. The final output will be calculated based on the states of the memory cell.

Consider time step \textit{t}, let $h_{t-1}$ and $c_{t-1}$ be previous hidden and cell state of LSTM layer, then we can compute current hidden state $h_{t}$, cell state $c_{t}$ can be computed by the following equations:
\begin{flalign}
i_{t}&=\sigma \left(U_{i}x_{t}+W_{i}h_{t-1}+b_{i}\right)\\
f_{t}&=\sigma \left(U_{f}x_{t}+W_{f}h_{t-1}+b_{f}\right)\\
o_{t}&=\sigma \left(U_{o}x_{t}+W_{o}h_{t-1}+b_{o}\right)\\
g_{t}&=\tanh \left(U_{g}x_{t}+W_{g}h_{t-1}+b_{g}\right)\\
c_{t}&=f_{t}c_{t-1}+i_{t}g_{t}\\
h_{t}&=o_{t}\tanh (c_{t})
\end{flalign}
where $\upsigma $ is sigmoid activation function, $U_{i}$, $U_{f}$, $U_{o}$, $U_{g}\in \mathbb{R}^{N\times d}$, $W_{i}$, $W_{o}$, $W_{f}$, $W_{g}\in \mathbb{R}^{N\times N}$, $b_{i}$, $b_{o}$, $b_{f}$, $b_{g}\in \mathbb{R}^{N}$ are learning parameters for LSTM, and $h_{t}$ is the output of LSTM cell. Note that \textit{N} is hidden layer size and \textit{d} is the dimension of the input vector. 

However, the standard LSTM network processes input in unidirectional order, which can make use of past context, but it ignores future information. To solve this problem, Bidirectional LSTM (BLSTM) introduces another hidden layer of opposite direction so that the output layer can exploit data from past and future concurrently\cite{schuster1997bidirectional,graves2005bidirectional}.

As shown in Figure 4, there are two sub-LSTMs pass forward and backward simultaneously. For time step \textit{t}, $h_{t}^{\left(l\right)}$denotes the output of the forward LSTM while $h_{t}^{\left(r\right)}$ denotes the output of the backward LSTM by reversing the order of word sequence. The output of the time step \textit{t} is:
\begin{equation}
h_{t}=(h_{t}^{\left(l\right)}\oplus h_{t}^{\left(r\right)})
\end{equation}
where $\oplus $ is element-wise sum.

\textit{3) Attention Layer:} Recently, attention mechanism has been proved effective in many NLP fields such as neural machine translation\cite{bahdanau2014neural,jiang2017automatically} and document classification\cite{yang2016hierarchical}. Specifically, attention mechanism is a model which can select more important parts in predicting the output label. For instance, the word ‘crash’ may be more discriminative in deciding whether an issue is a bug.

We introduced the attention mechanism into our architecture. Let \textit{H} be the matrix of outputs from Bi-LSTM. Namely, $H=[h_{1},h_{2},\ldots ,h_{m}]$, where \textit{m} is the sentence length. Then the output of attention layer is:
\begin{flalign}
M&=\tanh (H)\\
\alpha &=\textit{softmax}(w^{T}M)\\
r&=H\alpha ^{T}
\end{flalign}
where $w\in \mathbb{R}^{N}$ is a parameter vector to be learned, $\alpha \in \mathbb{R}^{m}$ is the attention weight and $r\in \mathbb{R}^{N}$. For every sentence, we calculate the final representation used for classification:
\begin{equation}
h^{*}=\tanh (r)
\end{equation}

\textit{4) Classification:} For a sentence \textit{S}, we predict label $\hat{y}$ by:
\begin{flalign}
p(y|S)&=\textit{softmax}(W^{0}h^{*}+b^{0})\\
\hat{y}&=\arg \max _{y} p(y|S)
\end{flalign}
where $W^{0}$ and $b^{0}$ are learning parameters.

\section{Experimental Design}\label{sec:experimental:design}

We conducted experiments on our approach to evaluate its effectiveness and efficiency. In this section, we describe the design of the experiments. 

We aim to address the following research questions:
\begin{itemize}
\item RQ1: Will our model predict better when misclassification corrector is enabled?

\item RQ2: Compared to the performance of the baseline methods and other similar approaches in this field, can we achieve a better performance?

\end{itemize}

\subsection{Dataset Detail}\label{sec:dataset}

We downloaded all of the issues in Apache ITS, JBoss ITS and Spring ITS (as of March 2019), all of which are based on Jira ITS. Note that a Jira ITS is typically configured to support multiple projects managed by an organization. Specifically, the Apache Jira ITS tracks the issues of 620 projects, including Zookeeper, Groovy, Hadoop, Maven, Struts and many other well-known Apache projects; 418 projects, including those for many JBoss components like JBoss Web, JBoss Reliance, and Netty, are hosted on the JBoss Jira ITS; Spring Jira tracks issues of 95 projects such as Spring Framework and Spring IDE. %All data are up to date in March 2019.

We also crawled 23 famous projects on GitHub, as described in Section~\ref{sec:data:collection}, to collect all their issues. We found it is harder to collect issues from GitHub than from Jira because only a few GitHub projects are well-labelled. We did not use an automatic approach to crawl all top-rank projects in GitHub because poorly labeled issues in many projects may have a negative impact on the discrimination power of our model. The GitHub projects we crawled are: AxonFramework, TypeScript, visualfsharp, vscode, OpenRefine, PowerShell, pulsar, deeplearning4j, che, elasticsearch, guava, google-cloud-java, hazelcast, javaparser, junit5, lettuce-core, micronaut-core, pinpoint, realm-java, spring-boot, spring-framework, spring-security, vavr. Table I reports some basic statistics of the issues we collected from various ITSs.

\begin{table}
\centering
\caption{Statistics of the issues from various ITSs.}
\begin{tabular}{p{0.7cm}p{0.7cm}p{1.7cm}p{1.2cm}p{1.2cm}}
\toprule
\multicolumn{2}{c}{\textbf{Data Source}} & \textbf{Labeled Issue Number}
 & \textbf{Project Number} & \textbf{Percent of Bugs}  \\ \midrule
 \multirow{3}{*}{Jira}  & Apache &  \makebox[1.3cm][r]{815338} & \makebox[0.7cm][r]{620} & 54.2\% \\
 & JBoss & \makebox[1.3cm][r]{329552} & \makebox[0.7cm][r]{418} & 49.3\% \\
 & Spring & \makebox[1.3cm][r]{65446} & \makebox[0.7cm][r]{95} & 39.7\% \\ \midrule
\multicolumn{2}{l}{GitHub} & \makebox[1.3cm][r]{85833} & \makebox[0.7cm][r]{23} & 44.7\% \\ \midrule
\multicolumn{2}{l}{Total} & \makebox[1.3cm][r]{1296169} & \makebox[0.7cm][r]{1156} & 51.6\% \\ \bottomrule
\end{tabular}
\end{table}

\subsection{Model Training and Testing Detail}

We based our model on Tensorflow\cite{abadi2016tensorflow}, an open source software which implements the underlying framework of neural network training. Tensorflow is user-friendly, robust and has been proven reliable in deep learning software development. 

Since there is no other RNN-based issue classification work, we borrowed and combined the training settings from the best-practice of RNN-based text classification work in other areas and fine-tuned the hyperparameters. The size of the word embeddings is 256; the size of the hidden layers is 256; the size of batches is 1024. To prevent over-fitting, we used dropout\cite{srivastava2014dropout} and set the dropout rate to 0.5. The model is validated every epoch in prediction accuracy. During the training, the model is saved every epoch. The maximum number of epochs is 20.

In testing phase, we loaded three models with highest validation accuracy and then obtained test accuracy respectively. After the training, we selected the model with the highest test accuracy for evaluation. What’s more, we also collected prediction details for every single issue in the test dataset and then calculated precision, recall, and F-measure. Ten-fold validation was also employed in the testing phase.

\subsection{Evaluation Metrics: Precision, Recall, and F-measure}\label{sec:metrics}

\emph{Precision} is defined as the number of true positive results divided by the total number of positive results predicted by a classifier, while \emph{recall} is the number of true positive results divided by the number of all positive results that should be assigned to positive. That is, recall measures ‘how complete the results are’ and precision measures ‘how error-free the results are’. \emph{F-measure} (also known as F-score) combines both the precision measure and the recall measure, and is calculated as the harmonic mean of precision and recall:
\begin{equation}
F-score=\frac{2\times \textit{precision}\times \textit{recall}}{\textit{precision}+\textit{recall}}
\end{equation}

We also calculate the weighted average value of F-measure $f_{avg}$ as the following for both classes in order to evaluate the overall performance, as was done in~\cite{fan2017road}: 
\begin{equation}
f_{avg}=\frac{n_{bug}\times f_{bug}+n_{\textit{nonbug}}\times f_{\textit{nonbug}}}{n_{bug}+n_{\textit{nonbug}}}
\end{equation}
where $f_{bug}$ and $f_{\textit{nonbug}}$ denote the F-measure for bugs and non-bug issues, respectively, while $n_{bug}$ and $n_{\textit{nonbug}}$ denote the numbers of bugs and non-bug issues, respectively.

\subsection{Evaluation for Misclassification Corrector\label{sec:determine-kp}}

In order to evaluate the effectiveness of our misclassification corrector, we randomly selected 3,000 labelled issues from our dataset and manually classified them to two categories: misclassified or correctly-classified. We recruited three postgraduate students for the task. All the students are majored in software engineering and experienced in open source software development. They are asked to first read and classify each selected issue independently, and then discuss the issues to which they assigned different types. An issue is only marked as being misclassified if all the students reach a consensus that the issue has a type different from its reporter type. % For the clarity of the judging standard, we defined misclassified issue as “an issue which is more possibly classified to the opposite of its present label”. We randomly selected 3,000 labelled issues in Jira and GitHub ITS. Three reporters read and annotated all of them respectively. Then they thoroughly discussed controversial ones.

For the revised k-NN algorithm used in our misclassification corrector, two parameters need tuning: the number of nearest neighbors \textit{k} and judgment threshold $p$. In a pilot study, we found that because our data set is large enough, the model has similar performance when \textit{k} varies in interval $[15,30]$. When \textit{k} is less than 15 or more than 30, the performance decreases significantly. Therefore, we chose a medium value \textit{k}=20 for a good balance between cost and effectiveness. For judgment threshold $p$, we conducted our experiments with $p$ = 1.0, 0.95, {\ldots}, 0.5, which means that the number of nearest neighbors of the different type should not be smaller than 20, 19, {\ldots}, 10 for an issue to be marked as misclassified. We use $M_{p}$ to denote the set of issues that are regarded as misclassified by our corrector with judgement threshold $p$. %Under these different circumstances, we collected the issues that are collected and corrected by the misclassification corrector. We use $M_{p}$ to denote the set of these issues.

Let $S$ be the set of 3000 issues we manually examined and $M$ the set of misclassified issues discovered in our manual analysis. We can then estimate the precision and recall of our misclassification corrector, w.r.t. a threshold $p$, using the following formulas:
%Then, we used our manually labelled data to extrapolate the measure of the whole dataset. Let's say $M$ is the set of misclassified issues in our manually labelled set, and $S$ represents our whole manually labelled set. Then, For threshold $p$, we calculated precision and recall by the following formulas:
\begin{flalign}
precision_{p} &=\frac{\left|M\cap M_{p}\right|}{\left|S\cap M_{p}\right|}\\
recall_{p} &=\frac{\left|M\cap M_{p}\right|}{\left|M\right|}
\end{flalign}

To measure to what extent misclassification corrector helps to improve the overall performance of bug classification, we first train a neural network model (described in Section~\ref{sec:classifying:issues}) with misclassification correction disabled, and we use the resultant model $M_{0}$ as the control group. Then, for each judgment threshold $p$ ($p\in\{ 1.0, 0.95, \ldots, 0.5\}$), we train another neural network model with misclassifications correction enabled. We use the same training set, validation set, and test set (these three sets are all applied with misclassification corrector) in all cases. At last, we calculated the precision, recall and F-score of the final classification.

\subsection{Baseline Selection and Implementation}

The bug classifier we propose aims at working effectively across multiple ITSs like Jira and GitHub. No existing approach, however, was designed with cross-platform applicability as a key distinguishing feature, and therefore can be used as baseline to evaluate our approach: Approaches proposed by Antoniol et al.~\cite{antoniol2008bug} and Pingclasai et al.~\cite{pingclasai2013classifying} utilize information like priority and severity for classifying issues, while issues on the GitHub ITS do not contain such information; The approach developed by Fan et al.~\cite{fan2017road} uses detailed developer information which is only available for issues in the GitHub ITS.

Therefore, we implement our own baseline bug classifiers using traditional machine learning algorithms like Logistic Regression (LR), Support Vector Machine (SVM), and k-Nearest Neighbor (k-NN). We intentionally included k-NN because it is also used in our misclassification corrector. Note however that the two implementations of k-NN are driven by different issue features and serve for different purposes. All the baseline approaches share the same data collection process as described in Section~\ref{sec:methodology}.
%that k-NN here serves as a classifier and is not the same approach as what is used in our corrector.

We use not only the whole dataset, but also partial datasets specific to the GitHub, Apache, JBoss, and Spring ITSs, to evaluate the performance of our approach and the baseline methods. We adopt the same experimental settings as used in~\cite{fan2017road}.
%Moreover, similar settings are used in baseline method implementation such as the same preprocessing technique (tokenization and stemming), and the same validation method (10-fold cross validation).

\section{Results}\label{sec:results}

\subsection{RQ1: Effect of Misclassification Corrector}\label{sec:eval:misclassification}

In this subsection, we first report the performance of misclassification corrector independently, and then we discuss the impact of the corrector on the whole model.

\textit{1) Performance of Misclassification Corrector with Variable Parameters:} 

Table II reports the precision, recall, and F-measure of our misclassification corrector. In general, we reasonably assume most issue types assigned by reporters are correct. Therefore, we try not to change those types unless there is strong evidence that they are wrong.
%Relatively, we valued precision more than recall because we prefer to avoiding reversing wrong issues and keeping large proportion of the dataset unaffected. 
We also estimated the correction rate in each case, which is the size of corrected data divided by the dataset size. The experiments indicate that by setting a judgment threshold larger than 0.5, the corrector successfully improves the precision of detection, without affecting large proportion of origin data. As we can see from the table, when $p$ increases, meaning our corrector selects and corrects items more 'strictly', the precision increases with $p$, recall decreases with $p$, and correction rate decreases with $p$.

\begin{table}[!tbh]
    \centering \setlength{\tabcolsep}{0.25em}\scriptsize
    \caption{Prediction Result of Misclassification Corrector under Different Judgment Threshold (C-Rate: Corrected-Rate)}
\begin{tabular}{lrrrrrrrrrrr}
\toprule
\textbf{p} & \textbf{1} & \textbf{0.95} & \textbf{0.9} & \textbf{0.85} & \textbf{0.8} & \textbf{0.75} & \textbf{0.7} & \textbf{0.65} & \textbf{0.6} & \textbf{0.55} & \textbf{0.5} \\
\midrule
Precision & 0.88  & 0.873 & 0.85  & 0.837 & 0.789 & 0.739 & 0.672 & 0.59  & 0.516 & 0.453 & 0.4 \\
\midrule
Recall & 0.018 & 0.06  & 0.127 & 0.218 & 0.317 & 0.433 & 0.546 & 0.643 & 0.732 & 0.816 & 0.889\\
\midrule
F-measure & 0.035 & 0.112 & 0.221 & 0.346 & 0.452 & 0.546 & 0.603 & 0.615 & 0.605 & 0.583 & 0.552\\
\midrule
C-rate & 0.003 & 0.01  & 0.021 & 0.036 & 0.056 & 0.082 & 0.114 & 0.152 & 0.199 & 0.252 & 0.312 \\
\bottomrule
\end{tabular}%    \caption{Caption}
    \label{tab:my_label}
\end{table}

\begin{table*}
\caption{EXPERIMENT DATA DETAIL FOR RQ2}
\centering
\setlength{\tabcolsep}{0.5em}
\begin{tabular}{lp{4.055em}rrrrrrrrrrrrrrr}
\toprule
\multicolumn{2}{c}{\multirow{2}[2]{*}{}} & \multicolumn{3}{c}{\textbf{GitHub}} & \multicolumn{3}{c}{\textbf{Apache}} & \multicolumn{3}{c}{\textbf{JBoss}} & \multicolumn{3}{c}{\textbf{Spring}} & \multicolumn{3}{c}{\textbf{All}} \\ \cmidrule(lr){3-5}\cmidrule(lr){6-8}\cmidrule(lr){9-11}\cmidrule(lr){12-14}\cmidrule(lr){15-17}
\multicolumn{2}{c}{} & \textit{\textbf{Perc.}} & \textit{\textbf{Rec.}} & \textit{\textbf{F-M}} & \textit{\textbf{Perc.}} & \textit{\textbf{Rec.}} & \textit{\textbf{F-M}} & \textit{\textbf{Perc.}} & \textit{\textbf{Rec.}} & \textit{\textbf{F-M}} & \textit{\textbf{Perc.}} & \textit{\textbf{Rec.}} & \textit{\textbf{F-M}} & \textit{\textbf{Perc.}} & \textit{\textbf{Rec.}} & \textit{\textbf{F-M}} \\
\midrule
\multicolumn{1}{l}{\multirow{3}[2]{*}{LR}} & Bug   & 0.693 & 0.607 & 0.647 & 0.721 & 0.765 & 0.743 & 0.743 & 0.73  & 0.736 & 0.732 & 0.594 & 0.656 & 0.721 & 0.74  & 0.73 \\
      & Nonbug & 0.708 & 0.78  & 0.742 & 0.7   & 0.649 & 0.674 & 0.747 & 0.758 & 0.752 & 0.711 & 0.863 & 0.815 & 0.719 & 0.699 & 0.709 \\
      & Average & 0.701 & 0.702 & 0.7   & 0.712 & 0.712 & 0.711 & 0.745 & 0.745 & 0.745 & 0.756 & 0.759 & 0.753 & 0.72  & 0.72  & 0.72 \\
\midrule
\multicolumn{1}{l}{\multirow{3}[2]{*}{SVM}} & Bug   & 0.714 & 0.604 & 0.655 & 0.762 & 0.75  & 0.756 & 0.793 & 0.729 & 0.759 & 0.749 & 0.592 & 0.662 & 0.771 & 0.723 & 0.747 \\
      & Nonbug & 0.714 & 0.804 & 0.756 & 0.71  & 0.722 & 0.716 & 0.756 & 0.815 & 0.784 & 0.711 & 0.874 & 0.819 & 0.72  & 0.769 & 0.744 \\
      & Average & 0.714 & 0.714 & 0.711 & 0.738 & 0.737 & 0.738 & 0.774 & 0.772 & 0.772 & 0.762 & 0.764 & 0.758 & 0.747 & 0.745 & 0.745 \\
\midrule
\multicolumn{1}{l}{\multirow{3}[2]{*}{kNN}} & Bug   & 0.64  & 0.45  & 0.528 & 0.726 & 0.614 & 0.665 & 0.761 & 0.55  & 0.639 & 0.686 & 0.446 & 0.54  & 0.722 & 0.564 & 0.633 \\
      & Nonbug & 0.644 & 0.797 & 0.712 & 0.611 & 0.723 & 0.662 & 0.657 & 0.833 & 0.735 & 0.705 & 0.867 & 0.777 & 0.625 & 0.77  & 0.69 \\
      & Average & 0.642 & 0.643 & 0.63  & 0.673 & 0.664 & 0.664 & 0.708 & 0.694 & 0.687 & 0.697 & 0.7   & 0.684 & 0.675 & 0.664 & 0.661 \\
\midrule
\multicolumn{1}{l}{\multirow{3}[2]{*}{Our work}} & Bug   & 0.779 & 0.826 & 0.802 & 0.855 & 0.864 & 0.859 & 0.859 & 0.877 & 0.868 & 0.808 & 0.823 & 0.816 & 0.842  & 0.875  & 0.858 \\
      & Nonbug & 0.844 & 0.801 & 0.822 & 0.850 & 0.840 & 0.845 & 0.887 & 0.870 & 0.879  & 0.883 & 0.872 & 0.878 & 0.871 & 0.838 & 0.854 \\
      & Average & \textbf{0.814} & \textbf{0.812} & \textbf{0.813} & \textbf{0.853} & \textbf{0.853} & \textbf{0.853} & \textbf{0.874} & \textbf{0.874} & \textbf{0.874} & \textbf{0.853} & \textbf{0.853} & \textbf{0.853} & \textbf{0.857} & \textbf{0.856} & \textbf{0.856} \\
\bottomrule
\end{tabular}%

\end{table*}

\textit{2) Impact of Misclassification Corrector:} Figure 5 shows the trend of average F-measure changing with judgment threshold $p$. The orange line shows the results of the control group $M_{0}$. The average F-measure in $M_{0}$ is 0.843. In other words, if there is not a misclassification corrector, we can achieve an F-measure of 0.843. The blue line represents the performance curve of our model with the assistance of the misclassification corrector. The figure shows that our approach with misclassification corrector outperforms the control group when $p$ equals or more than 0.65. When $p$ is less than 0.65, performance decreases quickly. Also, $p$=0.8 is the optimal threshold. When $p$ equals 0.8, we can achieve an F-measure of 0.856. Considering the possible error of sampling, we think [0.75,0.85] is the reasonable interval of $p$. 

\begin{figure}
\centering
\includegraphics[width=0.48\textwidth]{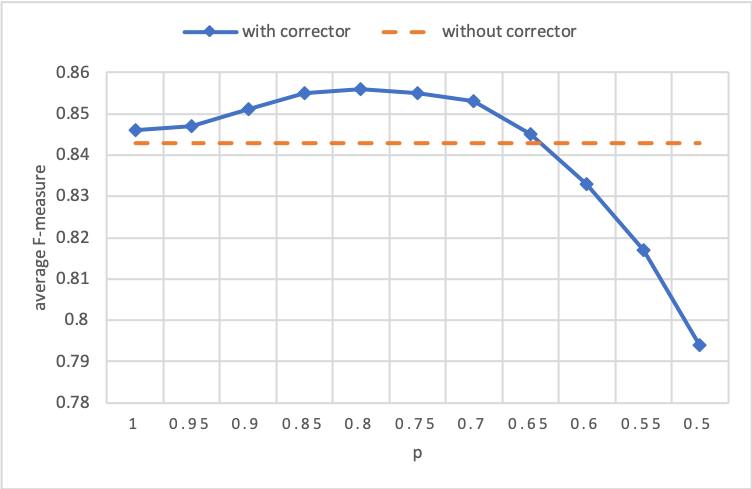}
\caption{F-score curve regarding judgment threshold $p$}
\end{figure}

%The last row of Table III shows the average F-measure when we applied the misclassification corrector to both the training set and the test set. The fact that F-measure increases quickly when $p$ increases implies that misclassified and ambiguous issues in test dataset are a major liability to achieve more accurate issue classification. Because we correct issues that are ambiguous or wrongly labelled in the test data set, the model is able to give a more convincing result. The findings indicate that although our recurrent neural network can significantly reduce the impacts of misclassification problem when building models providing large amount of training data in the dataset, misclassification still, to some extent, affects the evaluation of the model. 

\subsection{RQ2: Comparison with Baseline}\label{sec:eval:comparison}

\textit{1) Comparison with our baseline methods:} Figure 6 compares our result with baseline methods under different sub-dataset and the entire dataset. Table III provides details in our experiment. From the table we can see that our approach significantly outperforms all of baseline works under all datasets in terms of the weighted average F-measure, precision and recall. 

\begin{figure}
\centering
\includegraphics[width=0.48\textwidth]{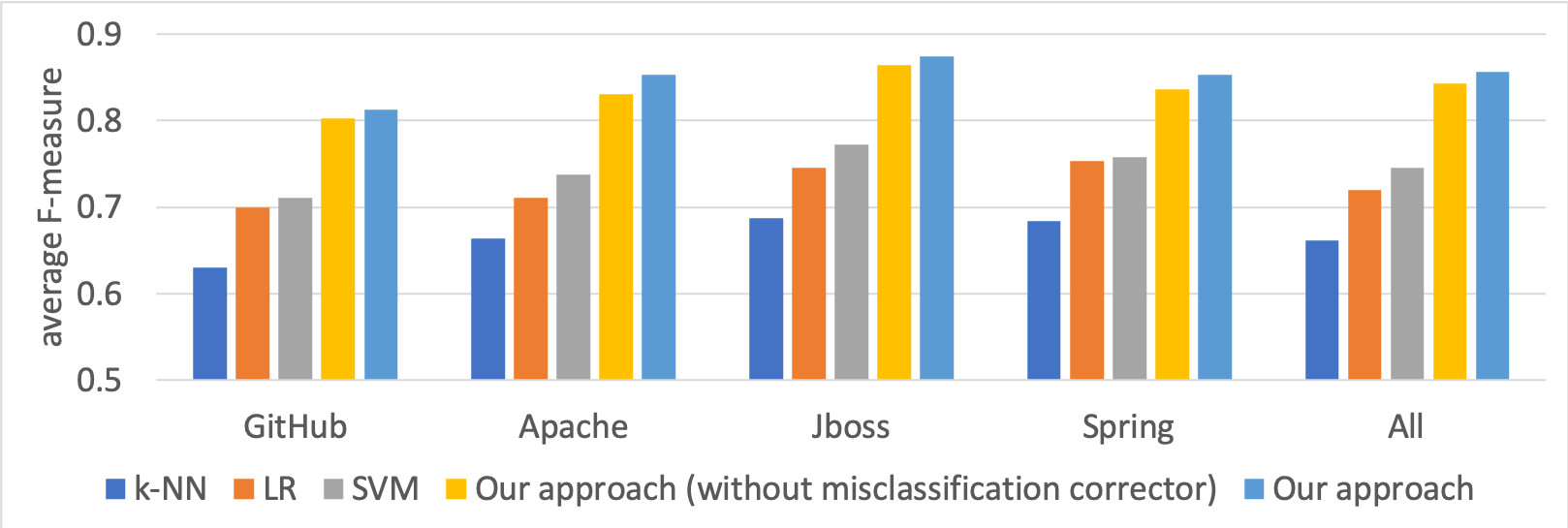}
\caption{Comparison with baseline works}
\end{figure}

Figure 6 shows that all of our classifiers are much better than traditional methods, and SVM performs better than k-NN and LR slightly, which echoes Fan et al.’s investigation\cite{fan2017road}. Specifically, compared with SVM, the best baseline method, our work has improved the classification weighted average F-measure from 71.1\% to 81.3\% for GitHub, from 73.8\% to 85.3\% for Apache, 77.2\% to 87.4\% for JBoss and from 75.8\% to 85.3\% for Spring. For the whole dataset, we increase the performance from 74.5\% to 85.6\%. In addition, whether in bug detection or in non-bug categorization, our model surpasses any of the baseline methods in both precision and recall measurement by a large gap.

Besides, we calculated the F-measure of our approach without corrector and drew the result together in Figure 6. Under all circumstances, our misclassification corrector can improve the model's performance, which reinforces our answer to RQ1.

We noticed that all the baseline methods and our approach get an obviously worse performance for GitHub data than for Jira data. The reason may be that the project number of our GitHub dataset is too small and thus the dictionary relies largely on the terminology of certain projects. The unbalanced proportion inside GitHub data distribution may also contribute to the low quality of GitHub data.

\textit{2) Comparison with the state-of-the-art methods:} The most recent comparable works in this field were done by Antoniol et al.\cite{antoniol2008bug}, Fan et al.\cite{fan2017road} and Pingclasai et al.\cite{pingclasai2013classifying}. We did not repeat their work because our dataset is cross-ITS, which means we only used data fields shared by all ITSs: label and textual description. Unfortunately, most of previous works are ITS-specific (Some discussions will be presented in Section~\ref{sec:relatedwork}) and thus cannot be repeated onto our dataset. In other words, these ITS-specific works needs extra data fields which are absent in our current dataset. 

Also, most previous researches were based on relatively small dataset, but our approach was designed to trained on large amount of issues. For instance, Zhou et al.\cite{zhou2016combining} trained on a dataset of only about 3k bug reports, involving 5 open source projects in Bugzilla and Mantis. Another example is the dataset provided by Herzig et al.\cite{herzig2013s}\footnote{https://www.st.cs.uni-saarland.de/softevo/bugclassify/} in 2013, which includes only 7401 issues from 5 JAVA projects.

Antonial et al.\cite{antoniol2008bug} distinguishes bugs from other kinds of issues, building their classifiers with between 77\% and 82\% of correct decisions, while the precision of our classifiers' bug prediction can reach 84.2\% and the precision of our classifiers non-bug prediction can reach 87.1\%. Moreover, we can achieve a higher recall. Fan et al.\cite{fan2017road} classified issues in GitHub. As they reported, their approach improved traditional SVM from 75\% to 78\% in F-measure, while our approach trained on GitHub dataset can improve F-measure from 71.1\% (for SVM) to 81.4\%. 

Pingclasai et al.\cite{pingclasai2013classifying} used topic modeling technique to distinguish bug report, involving about 5k issues on 3 open source projects. Instead of weighted average F-measure, they used micro F-measure to measure their model's performance. In Micro-average method, you need to sum up the individual true positives, false positives, and false negatives for both bug and non-bug issues and the apply them to get the statistics. They yields the micro F-measure between 0.65 and 0.82 for different projects. We also implemented micro F-measure and our approach reaches the micro F-measure of 85.6\% in overall dataset.

Therefore, although we did not use the same dataset and data preprocessing procedure as our precedents did, the results of our experiments strongly implied that the performance of our model is much better than those of other recent works in this field.

\section{Threats to Validity}\label{sec:threats}

We have identified several major threats to validity. The first threat is about the human annotation on misclassification. Different reviewers may have different standards or opinions about whether an issue is misclassified. Moreover, there are other different factors influencing misclassification annotation. For instance, Herzig et al.\cite{herzig2013s}, in 2013, found 33.8\% of all bug reports to be misclassified. In the same year, Wright et al.\cite{wright2013estimating} estimated that there are between 499 and 587 misclassified bugs in MySQL bug database, which includes 59447 bug reports in total. This shows the huge gap between different researches in identifying misclassification. From our manual labelling process, we roughly estimated about 10\%-15\% are misclassified in our database.

The second threat is the dataset we used. Because of the poor average issue label quality, we selected 23 projects in GitHub and collected over 170k issues. However, it may be too small compared to the size of the Jira data collected. In addition, there are other traditional ITSs which have not been included in our dataset, such as Bugzilla and Redmine. Although our approach has been qualified to handle issues regardless of which platform they belong to, further adjustment and evaluation are needed regarding other ITSs.

%Another threat is the performance of our misclassification corrector. The misclassification corrector only increases the model’s weighted average F-measure by 0.3\%, which may be too small and subject to statistical error. Therefore, we carried out a 10-fold cross validation to reduce the error. What’s more, our way to identify misclassification may help to improve the quality of data set in a explicable way. Nevertheless, our model greatly outperforms the state-of-the-art approach with or without the corrector.

\section{Related Work}\label{sec:relatedwork}

For space reasons, this section reviews researches that are the most relevant to this work in issue classification.
%At present, researches on Issue Tracking System are still developing, and many investigations have been carried out in issue mining field.

Antoniol et al.~\cite{antoniol2008bug} were among the first to research on the issue classification problem. In their approach, features were extracted from issue titles, descriptions, and related discussions, and traditional machine learning algorithms, such as Na\"{i}ve Bayes classifier and decision tree, were employed to classify issues. Given an issue, a considerable amount of discussions, however, may take place several days after the issue is reported, which may have a negative impact on the effectiveness of the approach when prompt classification of issues is expected.
Pingclasai et al.\cite{pingclasai2013classifying} apply topic modeling to the corpora of bug reports with traditional machine learning techniques including naive Bayes classifier, logistic regression and decision tree. Similar to\cite{antoniol2008bug}, they extract three contents of textual information: title, description and discussion. Their performance in classification, measured in F-measure, varies between 0.66-0.76, 0.65-0.77 and 0.71-0.82 for HTTPClient, Jackrabbit and Lucene project respectively.

Fan et al.~\cite{fan2017road} proposed an approach to classifying issues in GitHub. In their approach, features are extracted from both textual information of issues (including, e.g., issue title and description) and personal information of issue reporters, assuming that the background of the reporters may influence classification. For example, they thinks skilled developers are likely to report a bug-prone issue and provide more useful bug reports. The median of weighted average F-score for the approach was around 0.78, while the median F-score from using SVM is about 0.75, suggesting that ITS-specific data can be utilized to achieve better classification results. In comparison, our approach uses data that are easier to collect and it can achieve better F-measure.

Compared to other issue classification works, the work done by Zhou et al.~\cite{zhou2016combining} is special because they did not try to predict type out of a raw issue, but aimed at answering the question of whether a given bug-labelled issue is a corrective bug description or only documenting developers' other concern. They utilized structural information, including priority and severity, of issues that are available in most ITSs. But in lightweight ITSs like the one used in GitHub, issues do not necessarily have such information.

%poses new challenges . It is not clear how well their approach performs when little only limited information can be drawn from  discussion data is available.

%Information from issue tracking systems has also been utilized to facilitate other activities in software development. For instance, Weiss et al.~\cite{weiss2007long} mined issues of the JBoss project to predict the average bug-fixing effort. Guo et al.~\cite{guo2011not} studied bug report reassignment and discovered that, contrary to common impression, reassignments are useful to determine the optimal person for bug fixing. They also made recommendations to improve the social-awareness of ITSs. Meneely et al.~\cite{meneely2010improving} extracted the relationship between developers and issues to understand and measure the project development artifacts and problem-solving process.

\section{Conclusion}\label{sec:conclusion}

In this paper, we proposed a novel approach to automatically distinguish bug and non-bug issues in Issue Tracking Systems. Our strategy was, in a nutshell, to 1) collect a large scale of data from different ITSs, 2) preprocess and correct misclassification issues that may harm the model's performance, and 3) train an attention-based bi-directional LSTM network to label issues with `bug' or `non-bug' tags. We carried out an empirical study, which shows that our approach outperforms the state-of-the-art approaches and achieves better results on text classification evaluation metric. Our approach is easier to apply across different ITSs, since it only requires issue titles as input.

% Our study shows the necessity of exploring the misclassification problem to improve the results, and moreover, provides a means to balance the manual effort and result accuracy. To correct misclassification issues sometimes require more manual effort, as developers are not sure which ones are misclassified. With our approach, developers can focus on the ones filtered out as “possible misclassifications”, and thus save the manual inspection effort.

% In the future, we will expand our dataset to include more GitHub data and data in other ITSs. We also aim to mine more information from issues, like extracting bug description from issues and grouping issues that describe the same bugs. At last, we plan to distinguish other types of issues such as feature request or enhancement suggestion.

\bibliographystyle{IEEEtran}
\bibliography{IEEEabrv,mylib}
\end{document}